\documentclass[aps,twocolumn]{revtex4}
\usepackage{adjustbox}
\usepackage{epsfig}
\usepackage{graphicx}
\usepackage{amsmath,amssymb,amsfonts}
\usepackage{array}
\usepackage{url}
\usepackage[colorlinks=true,linktocpage=true,linkcolor=blue,citecolor=blue]{hyperref}
\usepackage{multirow}
\usepackage{float}
\usepackage{lineno}
\usepackage{xspace}
\usepackage{subcaption}
\usepackage[usenames,dvipsnames]{color}
\usepackage{lineno,hyperref}
\usepackage{xspace}
\usepackage{multirow}
\usepackage{ragged2e}
\newcommand{\sqsn}{$\sqrt{s_{_{\rm{NN}}}}$\xspace}

\newcommand{\pt}{$p_{\rm{T}}$\xspace}

\newcommand{\coal}{$``$Coalescence''\xspace}
\newcommand{\trpt}{$``$Transport''\xspace}
\begin{document}

\title{Exploring light nuclei production at RHIC and LHC energies with A Multi-Phase Transport model and a coalescence afterburner}
\author{Yoshini Bailung$^{1}$} \thanks{yoshini.bailung.1@gmail.com}
\author{Neha Shah$^{2}$}
\thanks{nehashah@iitp.ac.in}
\author{Ankhi Roy$^{1}$}
\thanks{ankhi@iiti.ac.in}
\affiliation{$^{1}$Department of Physics, Indian Institute of Technology Indore, Simrol, Indore, Madhya Pradesh, India}
\affiliation{$^{2}$Department of Physics, Indian Institute of Technology Patna, Bihta, Patna, Bihar, India}

\begin{abstract}

In heavy-ion collisions, understanding how light nuclei species are produced can provide insight into the nature of hadronic interactions in extreme conditions. It can also shed light on understanding the matter-antimatter asymmetry and dark matter searches in astrophysical processes. To investigate the production mechanism of light nuclei such as deuteron, triton, and helium-3, we use a naive coalescence afterburner coupled to the well-known $``$A Multi-Phase Transport model" (AMPT). We focus on studying the production of light nuclei in central Au+Au collisions at different center of mass energies (\sqsn = 19.6, 39, and 200 GeV) and in Pb+Pb collisions at \sqsn = 2.76 TeV, at mid-rapidity. We generate events with the string melting version of AMPT, and feed the information of the nucleons with spatial and momentum conditions into the coalescence afterburner. Our study reports differential and integrated yields in transverse momentum (\pt) of the light nuclei in different center of mass energies. We also estimate the coalescence parameters ($B_A$) as a function of \pt and collision energy for (anti-)deuterons, tritons and helium-3s for Au+Au and Pb+Pb collisions, which are compared to other light nuclei production studies. All results are compared with measurements from the STAR and ALICE experiments.

\end{abstract}
\date{\today}
\maketitle

\section{Introduction}

\label{introduction}
Experiments at the Relativistic Heavy-Ion Collider (RHIC), BNL, USA~\cite{STAR:2005gfr} and the Large Hadron Collider (LHC), CERN, Geneva,~\cite{ALICE:2008ngc} are designed to conduct high energy hadronic and heavy-ion collisions and examine the ensuing produced particles. Amongst all, measurements of light (anti-)nuclei and hyper (anti-)nuclei production are of great importance to understand the nature of hadron-hadron interactions and extract the freeze-out parameters of the QCD matter produced in these collisions. These studies also lead towards unfolding the mysteries of matter-antimatter asymmetry and provide relevant insights in the search for signatures of dark matter~\cite{Winkler:2020ltd}. In hadronic or heavy-ion collision experiments, (anti-)nuclei production mechanisms are not well understood. They are believed to be produced in the final stages of the hadronic evolution, specifically during chemical and kinetic freeze-out. The typical binding energies of light nuclei are of $\mathcal{O}(2 \ \rm{MeV})$, which is much smaller than the temperatures during chemical and kinetic freeze-out $\mathcal{O}(100 \ \rm{MeV})$.\\

A substantial amount of studies with theoretical models have described how the light (anti-)nuclei and hyper (anti-)nuclei are produced in these extreme conditions. Models such as the thermal model, which consider the system to be in chemical and kinetic equilibrium, are successful in reproducing the anti-baryon over baryon and light anti-nuclei over nuclei ratios~\cite{Andronic:2010qu,Cleymans:2011pe,Steinheimer:2012tb}. On the other hand, coalescence models; based on the ansatz that the formation of light (anti-)nuclei and hyper (anti-)nuclei occur after the system has reached a kinetic freeze-out successfully describe nuclei production and the respective particle ratio measurements from experiments~\cite{Shah:2015oha,Zhao:2021dka}. In a coalescence approach, nucleons come close in phase space to form nuclei. The coalescence parameter ($B_{A}$) represents the probability of production via coalescing nucleons at kinetic freezeout, and is inversely related to the nucleon correlation volume (volume of nuclear matter during the time coalescence of nucleons to nuclei). Often the coalescence mechanism is applied as an afterburner, after being driven by transport or hydrodynamic model for phase space distribution of nucleons~\cite{Sun:2020uoj,Hillmann:2021zgj,Zhao:2020irc,Zhu:2015voa}. Moreover, kinetic processes or potential interactions during the hadronic evolution also serve as a tangible mechanism for nuclei production. Recent advancements in n-body transport models~\cite{Aichelin:1991xy}are allowing detailed and concrete studies in light as well as hyper nuclei cluster formation via potential interactions~\cite{Glassel:2021rod}.\\

The first measurements in light (anti-)nuclei production were carried out at the CERN Inner Storage Ring experiments~\cite{Alper:1973my,British-Scandinavian-MIT:1977tan}. Several measurements that followed in RHIC and the LHC reported light nuclei and hyper nuclei ($d,\bar{d}$, $\rm{^3H}$, $\rm{^4\overline{He}}$, $\rm{^4_{\Lambda}\overline{He}}$) production in $p+p$, Au+Au, $p$+Pb and Pb+Pb systems in different center of mass energies~\cite{STAR:2019sjh,ALICE:2015wav,ALICE:2015oer,STAR:2011eej,STAR:2022hbp}. All these studies open a lot of doors to test the model calculations. The dependence of nuclei production with energy allow researchers to narrow down the location of the critical point in the QCD phase diagram. The coalescence parameter ($B_{A}$) is reported in these measurements to have a finite slope as a function of \pt. This is regarded to as an increase in coalescence probability, or a decrease in the nucleon correlation volume with increasing \pt.\\

This article presents the implementation of a simple transport + coalescence methodology for producing light nuclei. The coalescence model, which requires only a small number of tuning parameters, effectively captures light nuclei ($d,\bar{d}$, $\rm{^3H}$, $\rm{^3He}$) production for a broad range of beam energies in heavy-ion collisions. In section~\ref{amptmodel}, the AMPT model~\cite{Lin:2004en} is described, with its use to obtain the phase space information of the constituent nucleons. We employ the string melting version of AMPT (version 1.26t9b-v2.26t9b), which is then coupled to a coalescence afterburner, from which the light nuclei yields are estimated. The coalescence model and its implementation is discussed in section~\ref{coalmodel}. AMPT also employs kinetic processes for deuteron and anti-deuteron formation during the hadronic transport, which gives an opportunity for a comparative study between the two modes (transport vs coalescence). The study is carried out in various center of mass energies \sqsn = 19.6, 39, and 200 GeV for central (0-10\%) Au+Au collisions and \sqsn = 2.76 TeV for central (0-10\% and 0-20\%) Pb+Pb collisions at mid-rapidity. We also compare our results with some recent predictions from the parton-hadron-quantum-molecular-dynamics (PHQMD) model~\cite{Aichelin:2019tnk} with light nuclei clustering algorithms~\cite{Glassel:2021rod}. The nuclei clustering is done dynamically in various cluster freezeout times via Simulated Annealing Clustering Algorithm (SACA) and Minimum Spanning Tree (MST) algorithm~\cite{Puri:1998te,Aichelin:2019tnk}. In section~\ref{results}, the results are presented and compared to the existing experimental measurements from the STAR and ALICE experiments.

\section{The AMPT model}
\label{amptmodel}
AMPT~\cite{Lin:2004en} consists of a hybrid framework of four primary components; the initial conditions, parton level interactions, a hadronization mechanism, and the hadronic scatterings. The initial conditions comprise the hard scatterings between mini-jet partons and soft scattering strings from the Heavy-Ion Jet INteraction Generator (HIJING) model~\cite{Wang:1991hta}. Zhang Parton Cascade (ZPC) model~\cite{Zhang:1997ej} handles the parton evolution and the scatterings between them, which recombine to form hadrons via the Lund string fragmentation model~\cite{Andersson:1983ia}. However, in the string melting~\cite{Lin:2001zk} version of AMPT, the ZPC handles both minijet partons and the excited strings during parton evolution, which then goes through hadronization via a parton coalescence mechanism~\cite{Molnar:2003ff}. The final stage is the hadronic evolution, which is brought in by A Relativistic Transport (ART) model~\cite{Li:1995pra} and additional resonance channels that contribute to the production of final state hadrons.\\

The AMPT model is well established in describing various experimental observables obtained at RHIC and the LHC~\cite{STAR:2010vob, ALICE:2013rdo, STAR:2013ayu}. Several interpretations of nuclei formation are covered using AMPT. AMPT natively produces (anti-)deuterons via elementary interactions of $NN \rightarrow d\pi$, $NNN \rightarrow d N$, and $NN\pi \rightarrow d\pi$ during the hadronic evolution. In this study, we investigate the deuteron production processes in two different modes. For \coal mode, we turn off the native deuteron processes in AMPT to ensure that all (anti-)nucleon pairs are available for the coalescence afterburner. For the \trpt mode, we explicitly turn on the hadronic rescattering processes and do not implement the afterburner on this freeze-out hypersurface.
\\
We generate AMPT events for the phase space information of (anti-)nucleons which are fed to the coalescence afterburner to produce (anti-)deuterons, tritons and helium-3s. We generate events for central Au+Au collisions at \sqsn = 19.6, 39, and 200 GeV and central Pb+Pb collisions at \sqsn = 2.76 TeV. For the Pb+Pb system, an additional parton mini-jet momentum cutoff at 5 GeV/$c$ is chosen which minimizes the softening of the (anti-)proton \pt spectrum. The ALICE results for Pb+Pb collisions at \sqsn = 2.76 TeV is well described with the application of an additional cut-off on the parton mini-jet momentum~\cite{Zhang:2019utb}. The Lund String Fragmentation function, which is defined as 
\begin{equation}
    f(z) \propto z^{-1} (1-z)^{a} \exp(-b m_{\perp}^{2}/z)
\end{equation}
where, $z$ denotes the light cone momentum fraction of a produced hadron with respect to that of the fragmentation string, $m_{\perp}$ is the transverse mass of the hadron, and $a$ and $b$ are the Lund Fragmentation parameters. It can be shown that the mean squared transverse momentum of the particles is proportional to the string tension ($k$) of fragmentation string ($\langle p_{\rm T}^{2}\rangle \propto 1 / b(2+a)$), which is related to the parameters $a$ and $b$ by
\begin{equation}
    k \propto \frac{1}{b(2+a)}
\end{equation}
The parameters ($a$ and $b$) which best describe the experimental data for protons AMPT are chosen. The same set of parameters are used for the phase space information of (anti-)nucleons, in order to coalesce them into the light nuclei species. The parton scattering cross-section is kept as 3$~mb$~\cite{Tiwari:2020ult}. The centrality of the collisions, 0-10\% and 0-20\% are taken into account by considering the collision impact parameter range within 0-4 $fm$ and 0-7 $fm$ respectively. A detailed description of $a$ and $b$ values for \sqsn = 19.6, 39, 200 GeV and 2.76 TeV is reported in Table~\ref{deuterontable}. A total of $\mathcal{O}(10^5)$ AMPT events are generated for each \sqsn = 19.6, 39, 200 GeV (Au+Au) and 2.76 TeV (Pb+Pb) corresponding to both centrality classes.\\

\begin{table}
\centering
\begin{tabular}{|c|c|c|}
\hline
$\sqrt{s_{\mathrm{NN}}}~(\rm{GeV})$ & $a$ & $b~(\rm{GeV^{-2}})$ \\
\hline
19.6 & 0.55 & 0.15 \\
39 & 0.55 & 0.15 \\
200 & 0.55 & 0.15 \\
2760 & 0.8 & 0.05 \\ \hline
\end{tabular}
\caption{\justifying Parameters ($a$ and $b$) for Lund String Fragmentation function chosen for Au+Au collisions at \sqsn = 19.6, 39, 200 GeV and Pb+Pb collisions \sqsn = 2.76 TeV corresponding to 0-10\% and 0-20\% centrality class.}
\label{deuterontable}
\end{table}
It is worth noting that the present iteration of the string melting AMPT has certain limitations in precisely characterizing baryons~\cite{He:2017tla}, a key aspect with regard to the findings presented in this paper. Notably, AMPT without transport deuterons (under)overestimates the yields of (anti-)protons at mid-rapidity. A possible explanation in this regard will be mentioned in Section~\ref{results}. Additionally, the model implements quark coalescence via separate conservation of the numbers of meson and (anti-)baryons per event. This constraint is removed with an upcoming string melting AMPT by assigning a control parameter for the quarks to form a baryon or a meson, during quark coalescence. Studies involving the development of an updated string melting AMPT with the improved quark coalescence prescription cites a better description of experimental measurements for baryons~\cite{He:2017tla}.

\begin{table}
\centering
\begin{tabular}{|l|l|l|}
\hline
Parameters                         & (Anti-)Deuteron & Triton/Helium-3 \\ \hline
$\Delta r_{max}~(fm)$    & 3.15     & 4.18            \\ \hline
$\Delta p_{max}$ (MeV/$c$) & 190      & 350             \\ \hline
\end{tabular}
\caption{\justifying Parameters ($\Delta p_{max}$ and $\Delta r_{max}$) used in coalescence of deuterons, tritons and helium-3.}
\label{paramtable}
\end{table}

\section{The Coalescence Model}
\label{coalmodel}
The process of light (anti-)nuclei and hyper (anti-)nuclei production in heavy-ion collisions can be explained by the coalescence mechanism of respective nucleons or baryons~\cite{Steinheimer:2012tb}. The model assumes that nuclei production begins at kinetic freeze-out during the later stages of a collision. In the limit of no-baryon number transport at mid-rapidity, the coalescence probability is described by the coalescence parameter $`B_{A}$', as

\begin{eqnarray}
    E_{A}\frac{d^3 N_{A}}{d^3p_{A}} &=& B_{A} \left(E_{p}\frac{d^3N_{p}}{d^3p_{p}} \right)^{Z}\left(E_{n}\frac{d^3N_{n}}{d^3p_{n}}\right)^{A-Z}\\
    &\approx& B_{A} \left(E_{p}\frac{d^3N_{p}}{d^3p_{p}} \right)^{A}\nonumber
    \label{b2formula}
\end{eqnarray}

where $E\frac{d^3 N}{d^3p}$ is the invariant yield, $A$ and $Z$ are the atomic mass number and the atomic number of (anti-)nucleons or (anti-)nuclei. To calculate $B_{A}$, the proton and light nuclei transverse momentum are related as $p_{\mathrm{T} A} = A \cdot p_{\mathrm{T} p}$. The coalescence conditions for two nucleons to form a $d,\bar{d}$, $\rm{^3H}$, $\rm{^3He}$ are characterized by their phase space distribution. In this case, we implement a box description of coalescence where we define conditions based on the relative distance $(\Delta r)$ and momentum $(\Delta p)$ of the nucleon pairs. The coalescence conditions are met when the relative momentum between the nucleons are less than a critical value ($\Delta p^{max}$), and the distance between the nucleons is less than twice the size of the nuclear force radius ($\Delta r^{max} < 2R_{0}$). $R_0$ is taken to be $\sim$1.6 fm for deuterons and $\sim$2.1 fm for triton and helium-3~\cite{Hillmann:2021zgj}. A summary of the coalescence parameters used for (anti-)deuteron, triton and helium-3 is tabulated in Table~\ref{paramtable}. A step by step procedure to achieve this phase space coalescence is described below:

\begin{enumerate}
    \item The (anti-)nucleon states, which is $pn(\bar{p}\bar{n})$ for $d(\bar{d})$, $pnn$ for $\rm{^3{H}}$ and $ppn$ for $\rm{^3{He}}$ are boosted to the center of mass frame. 
    \item If their relative momentum $\Delta p = \mid\vec{p}_{1} - \vec{p}_{2}\mid < \Delta p^{max}$ , their momenta are combined ($\vec{p}_{d} = \vec{p}_{p} + \vec{p}_{n}$ or $ \vec{p}_{\rm{^{3}H},\rm{^{3}He}} = \vec{p}_{p} + \vec{p}_{n,p} + \vec{p}_{n}$)
   
    \item Among the (anti-)nucleon pairs, the position(s) of the (anti-)nucleon(s) at an earlier time are reset to the (anti-)nucleon at the later freeze-out time. This is done to ensure that the pair (triplet) of nucleons do not interact between these two times.

    \item The relative distances between the pairs are checked so that the condition $\Delta r = \mid\vec{r}_{1} - \vec{r}_{2}\mid < \Delta r^{max}$ is satisfied.

    \item The spin-isospin coupling probabilities are taken into account (3/8 for a deuteron and 1/12 for triton and helium-3)~\cite{Hillmann:2021zgj,Xia:2014rua}. The finally chosen deuteron, triton and helium-3 candidates that fulfill the coalescence conditions are then removed from the phase space distribution. 
\end{enumerate}

\section{Result and Discussion}
\label{results}

 \begin{figure*}[!ht]
    \centering
    \includegraphics[scale=0.3]{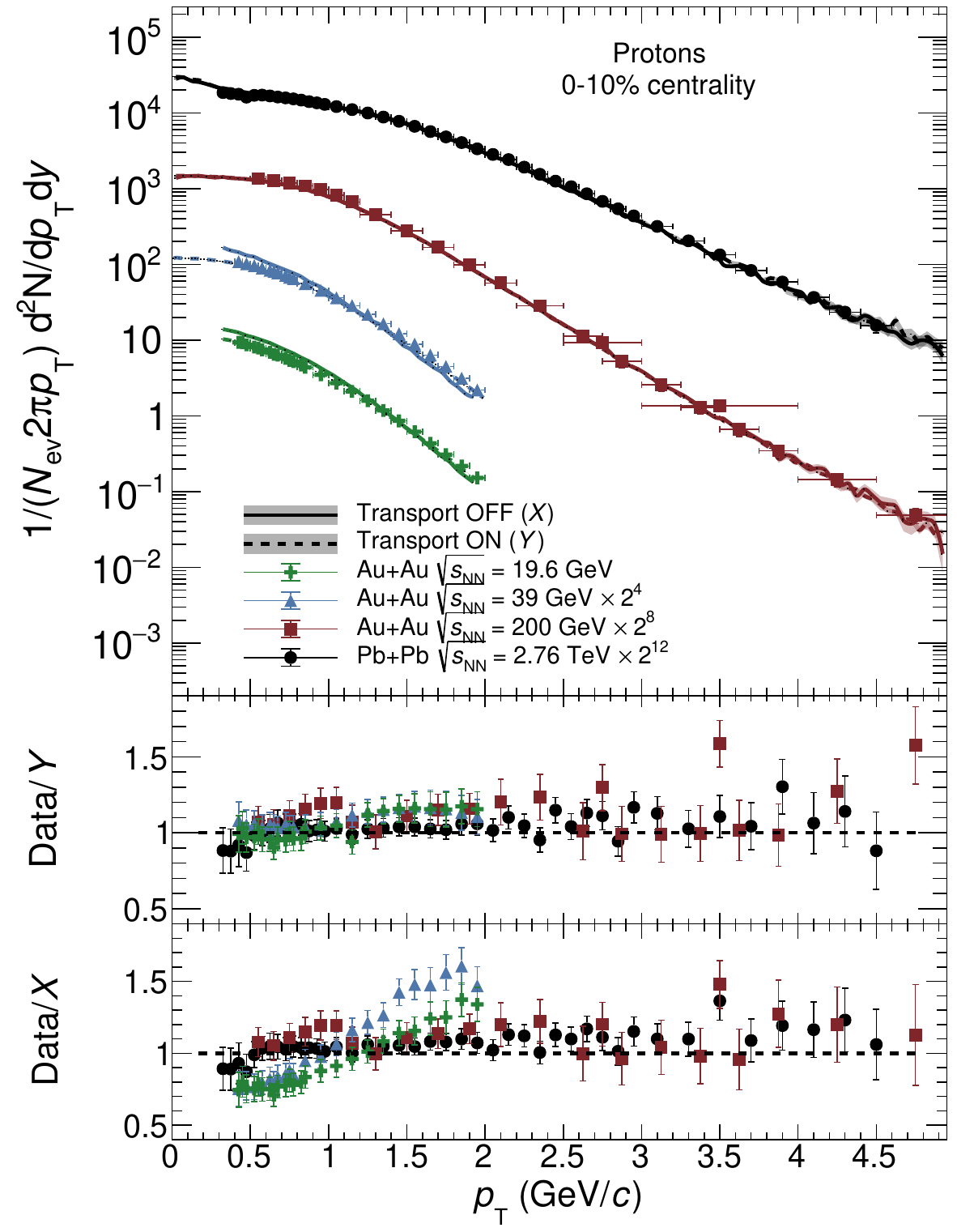}
    \hspace{1cm}
    \includegraphics[scale=0.3]{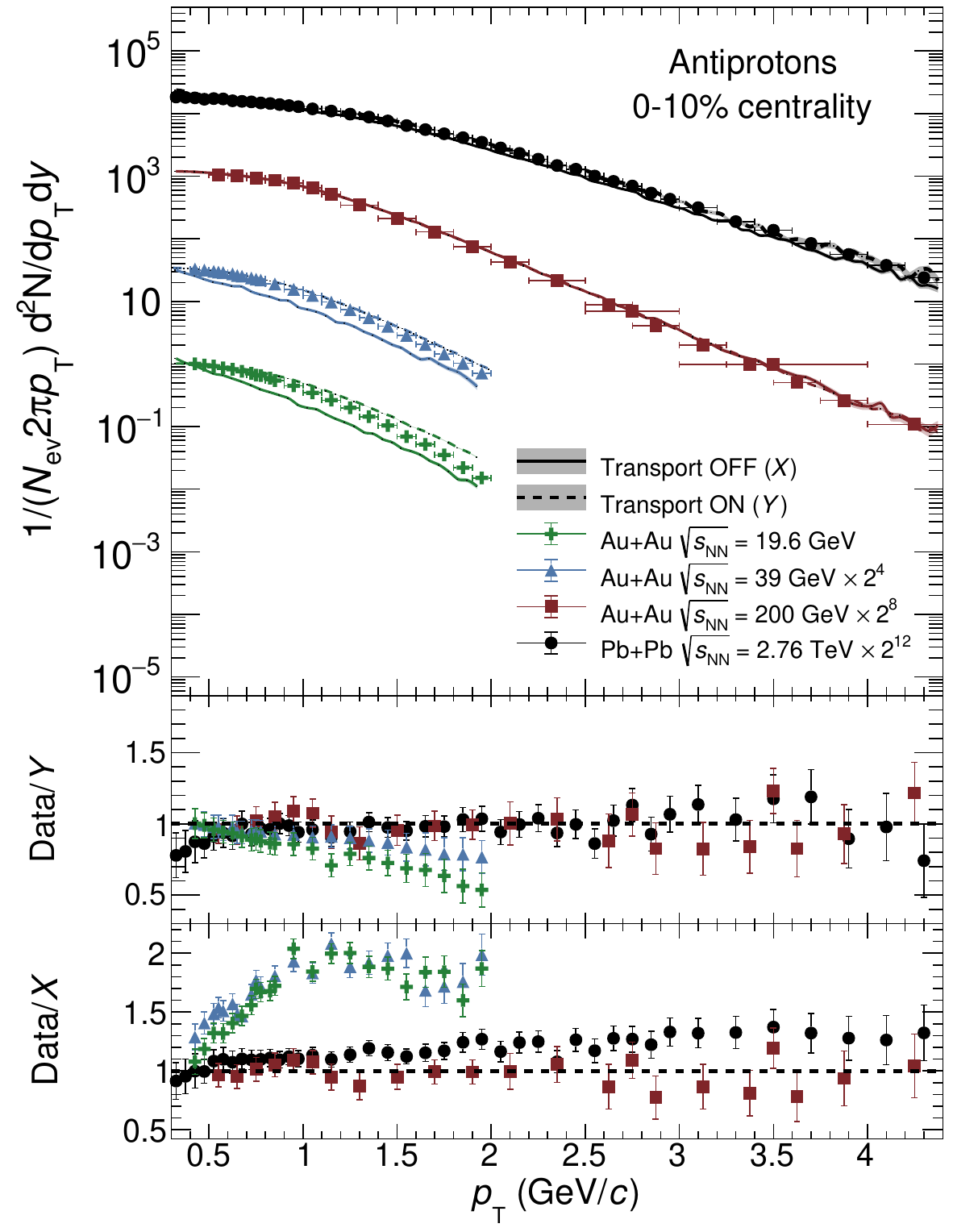}
    \caption{\justifying (Color online) Transverse momentum spectra of protons (left) and anti-protons (right) in central Au+Au collisions at \sqsn = 19.6 GeV, 39 GeV ($|y|<0.1$), 200 GeV ($|y|<0.5$) and central Pb+Pb collisions at \sqsn = 2.76 TeV ($|y|<0.5$) from \coal or $``$Transport OFF'' (solid lines) and $``$Transport ON'' (dashed lines) mode. Experimental results (symbols) are taken from STAR and ALICE experiments~\cite{STAR:2007zea,STAR:2017sal,ALICE:2013mez}. The lower panels show the ratio of the experimental results to the model predictions.}
    \label{proton}
\end{figure*}

\begin{figure*}[!ht]
    \centering
    \includegraphics[scale=0.3]{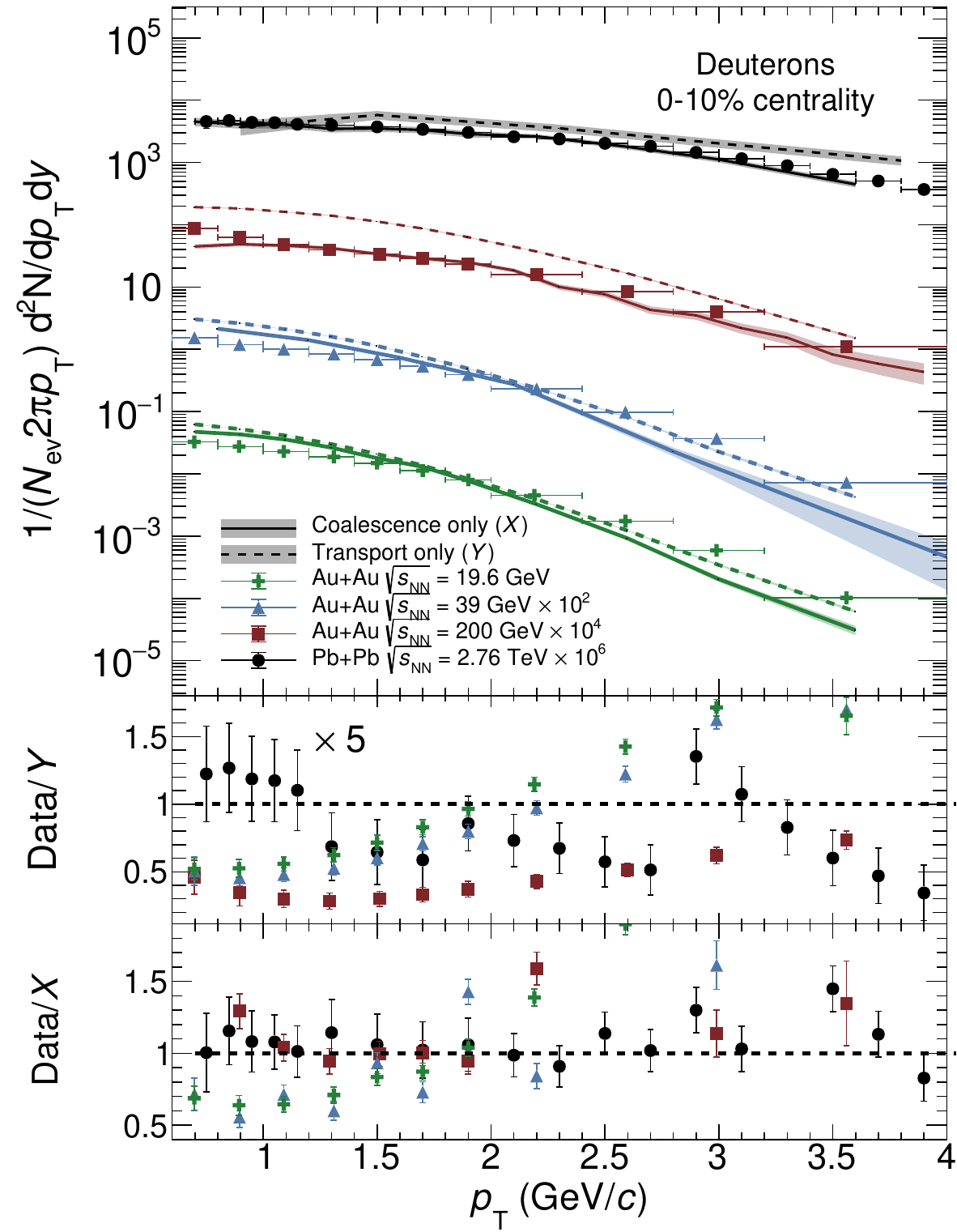}
    \hspace{1cm}
    \includegraphics[scale=0.3]{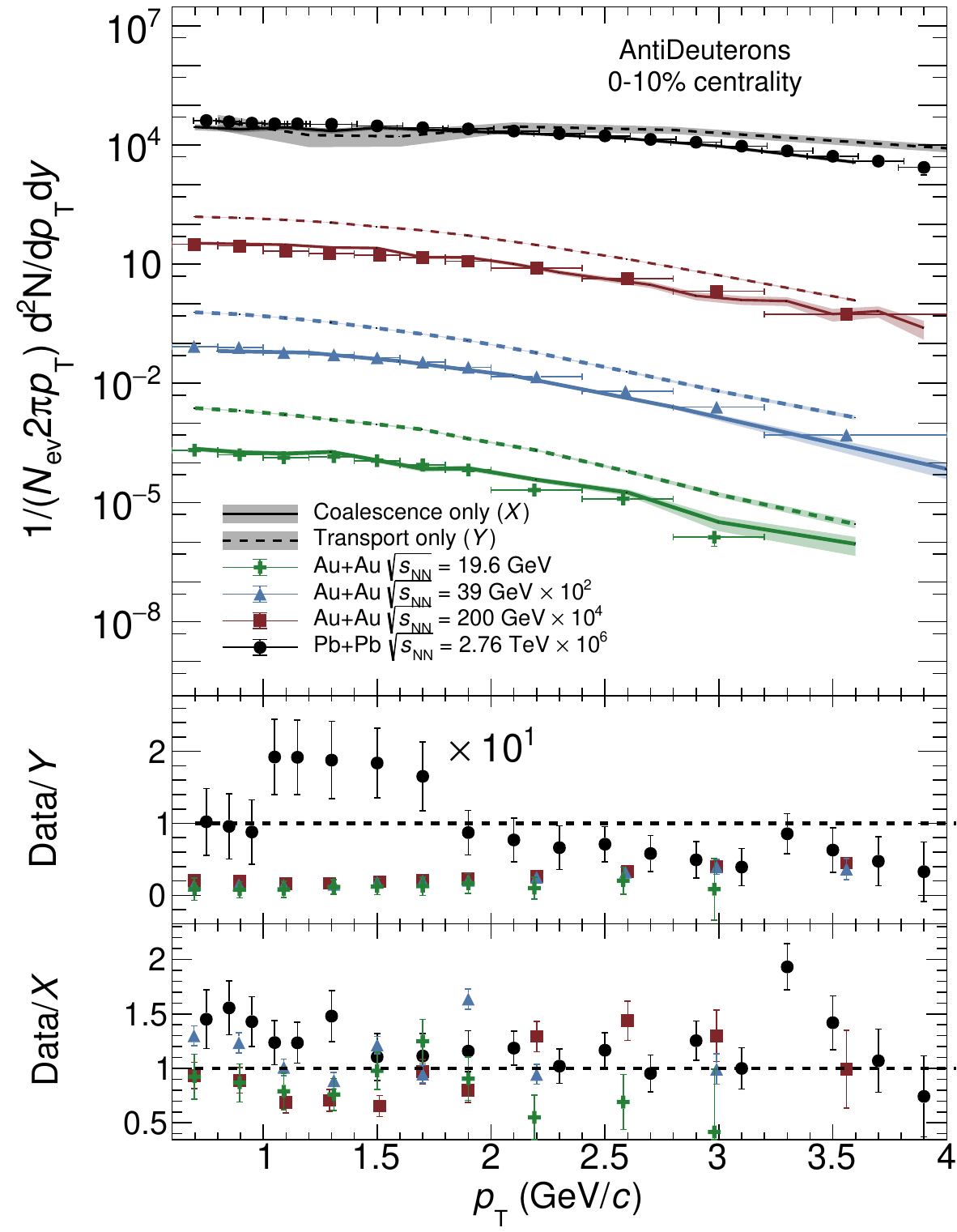}
    \caption{\justifying (Color online) Transverse momentum spectra of deuterons (left) and anti-deuterons (right) in central Au+Au collisions at \sqsn = 19.6 GeV, 39 GeV, 200 GeV ($|y| < 0.3$) and central Pb+Pb collisions at \sqsn = 2.76 TeV ($|y| < 0.5$) from \coal (solid lines) and \trpt (dashed lines) mode. Experimental results (symbols) are taken from the STAR and ALICE measurements~\cite{STAR:2019sjh,ALICE:2015wav}. The lower panels show the ratio of the experimental results to the model predictions.}
    \label{deuteron}
\end{figure*}

\begin{figure*}[!ht]
    \centering
    \includegraphics[scale=0.3]{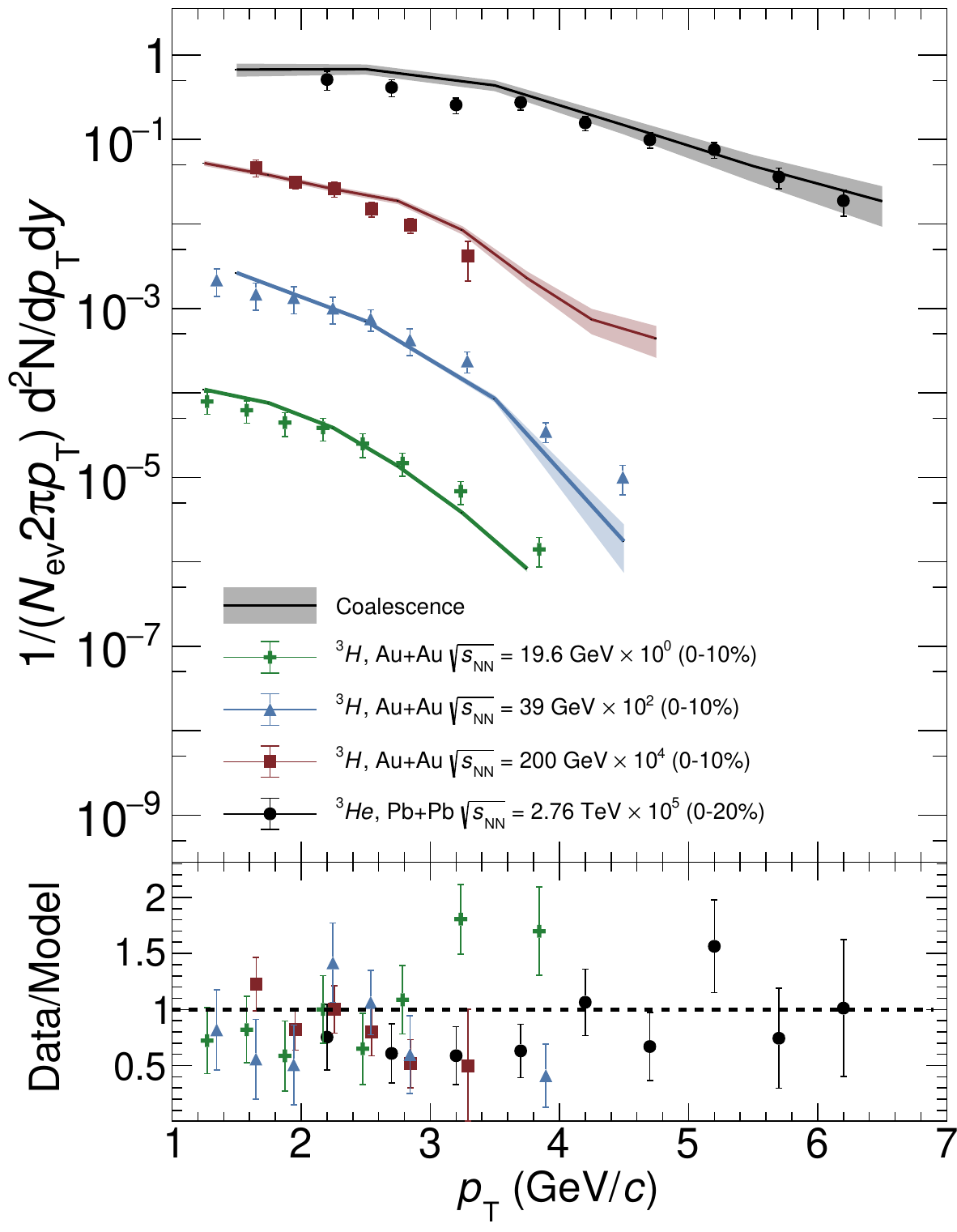}
    \caption{\justifying (Color online) Transverse momentum spectra of tritons in central Au+Au collisions at \sqsn = 19.6 GeV, 39 GeV, 200 GeV ($|y| < 0.5$) and helium-3s in central Pb+Pb collisions at \sqsn = 2.76 TeV ($|y| < 0.5$) from the \coal mode (solid lines). Experimental results (symbols) are taken from the STAR and ALICE measurements~\cite{STAR:2022hbp,ALICE:2015wav}. The lower panels show the ratio of the experimental results to the model predictions.}
    \label{trilion}
\end{figure*}

\begin{figure}[!ht]
    \centering
    \includegraphics[scale=0.45]{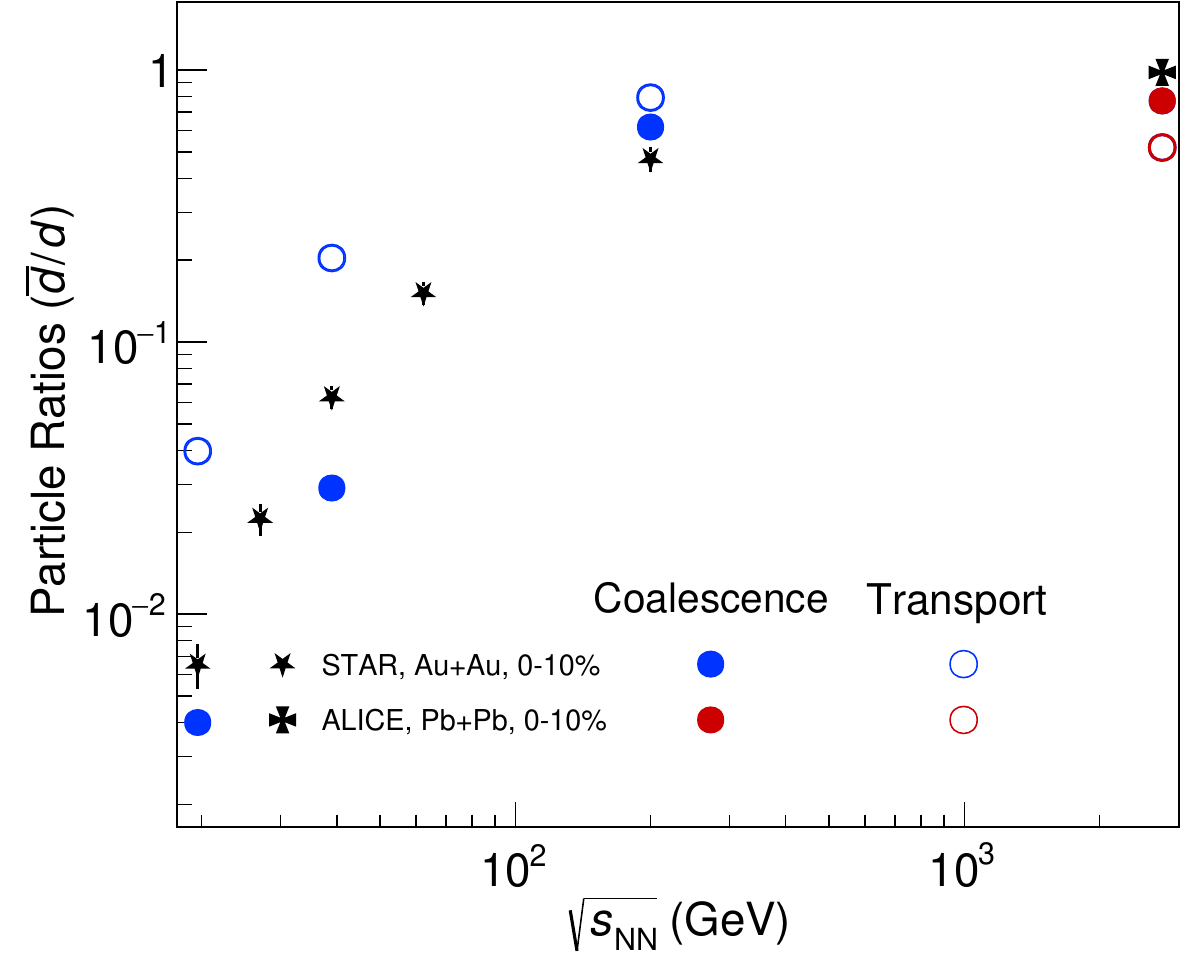}
    \caption{\justifying (Color Online) Particle ratios for anti-deuterons over deuterons as a function of center of mass energies from \coal and \trpt modes. Experimental measurment of $\bar{d}/d$ ratios from the STAR and ALICE experiments for the central Au+Au collisions at \sqsn = 19.6 GeV, 39 GeV, 200 GeV and central Pb+Pb collisions at \sqsn = 2.76 TeV are also shown~\cite{STAR:2019sjh}\cite{ALICE:2015wav}}
    \label{dbardratio}
\end{figure}

The transverse momentum spectra (\pt) for the (anti-)proton and their comparisons to the STAR and ALICE measurements~\cite{STAR:2007zea,STAR:2017sal,ALICE:2013mez} at mid-rapidity for \sqsn = 19.6, 39 ($|y| < 0.1$), 200 GeV ($|y| < 0.5$) (Au+Au, 0-10\%) and 2.76 TeV (Pb+Pb, 0-10\%, $|y| < 0.5$) respectively, are shown in Figure~\ref{proton}. The \coal (\textit{transport deuterons are OFF}) results overestimate the data for protons at \pt $< 1$ GeV/$c$ for \sqsn = 19.6 and 39 GeV and underestimate at \pt$\sim$ 2 GeV/$c$. In this aspect, the \trpt (\textit{transport deuterons are ON}) mode provides a better description of the \pt spectra for \sqsn = 19.6 and 39 GeV~\cite{STAR:2017sal}. Both the modes show good agreement with the data for protons at \sqsn = 200 GeV (Au+Au) and \sqsn = 2.76 TeV (Pb+Pb)~\cite{STAR:2007zea,ALICE:2013mez}. However, the \coal mode largely underestimates the data for Au+Au collisions at \sqsn = 19.6 and 39 GeV, but describe the experimental measurements at \sqsn = 200 GeV and 2.76 TeV. The anti-protons from \trpt mode provides a slightly harder \pt spectra for \sqsn = 19.6 and 39 GeV when compared to data, and performs similar to the \coal mode for \sqsn = 200 GeV and 2.76 TeV.\\

Figure~\ref{deuteron} displays the (anti-)deuteron \pt spectra for \coal and \trpt modes, which are compared to experimental measurements at midrapidity for Au+Au collisions at \sqsn = 19.6, 39, 200 GeV ($|y| < 0.3$), as well as Pb+Pb collisions at \sqsn = 2.76 TeV ($|y| < 0.5$). While the deuterons produced via \coal accurately reproduce the experimental data at \sqsn = 200 GeV (Au+Au)~\cite{STAR:2019sjh} and 2.76 TeV (Pb+Pb)~\cite{ALICE:2015wav} at all \pt values, they tend to overestimate the data at low \pt and underestimate it at high \pt in Au+Au collisions at \sqsn = 19.6 and 39 GeV~\cite{STAR:2019sjh}. This disagreement is attributed to the softer proton \pt spectra from AMPT. The \coal anti-deuteron spectra show minor deviations across \pt, and fit well with the experimental measurements for Au+Au and Pb+Pb collisions.~\cite{STAR:2019sjh, ALICE:2015wav}. In contrast, the deuterons and anti-deuteron yields from \trpt mode significantly deviates from the experimental results. In particular, the anti-deuteron yields are largely overestimated for Au+Au collisions. While, the \trpt deuteron spectra for Au+Au collision at \sqsn = 19.6 and 39 GeV sits close to the \coal spectra, the yields are overestimated for \sqsn = 200 GeV. In Pb+Pb collisions, the \trpt (anti-)deuteron yields are underestimated, with a visible drop at low \pt.\\

The results shown in Figure~\ref{proton} and \ref{deuteron} clearly demonstrate a trade-off between (anti-)proton and (anti-)deuteron production in the AMPT model. This trade-off likely contributes to the observed differences in (anti-)proton yields between the \coal and \trpt modes, especially for anti-proton yields in Au+Au collisions at \sqsn = 19.6 GeV and 39 GeV. The overestimation of anti-deuteron yields when using the \trpt mode suggests that there might be an inherent imbalance in the cross-sections governing anti-deuteron production and dissociation processes. This imbalance directly impacts the anti-proton yields when these processes are turned \textit{OFF}. In the \coal mode, where these kinetic processes are absent, the production of anti-protons is underestimated, particularly at collision energies where the net proton number is non-zero.\\

The triton and helium-3 \pt-spectra results from \coal mode are presented and compared to experimental results in Figure~\ref{trilion} for central (0-10\%) Au+Au collisions at \sqsn = 19.6, 39, 200 GeV~\cite{STAR:2022hbp} and central (0-20\%) Pb+Pb collisions at \sqsn = 2.76 TeV~\cite{ALICE:2015wav} at mid-rapidity ($|y| < 0.5$). For tritons in Au+Au collisions at \sqsn = 19.6 and 39 GeV, the model exhibits a deviation across \pt that follows a similar pattern to that observed for deuterons. The model slightly overestimates the experimental data at low \pt and underestimates at larger values. However, the behaviour changes for tritons in Au+Au collisions at \sqsn = 200 GeV, giving a harder spectra when compared to the data. The helium-3 spectra for Pb+Pb collisions at \sqsn = 2.76 TeV is well described, with a slight overestimation for \pt $< 3.5$ GeV/$c$.\\

The \pt-integrated yields for $d,\bar{d}$ and $\rm{^3H}$, $\rm{^3He}$ are calculated for central Au+Au collisions at \sqsn = 19.6, 39, 200 GeV and Pb+Pb collisions at \sqsn = 2.76 TeV at mid-rapidity via \coal and \trpt modes. Figure~\ref{dbardratio} shows the $\bar{d}/d$ ratios as a function of center of mass energy compared to experimental measurements. Both modes are able to qualitatively describe the trend of energy dependence of $d$ and $\bar{d}$ production successfully at the respective RHIC and LHC energies\cite{STAR:2019sjh}. Quantitatively, the \trpt mode systematically overestimates the experimental data for Au+Au collisions from 19.6 to 200 GeV, but largely underestimates the value for Pb+Pb collisions at \sqsn = 2.76 TeV. The \coal mode underestimates the ratios for Au+Au collisions at \sqsn = 19.6 and 39 GeV, and slightly overestimates for \sqsn = 200 GeV. The value at \sqsn = 2.76 TeV is slightly underestimated.\\
\begin{figure}[!ht]
    \centering
    \includegraphics[scale=0.45]{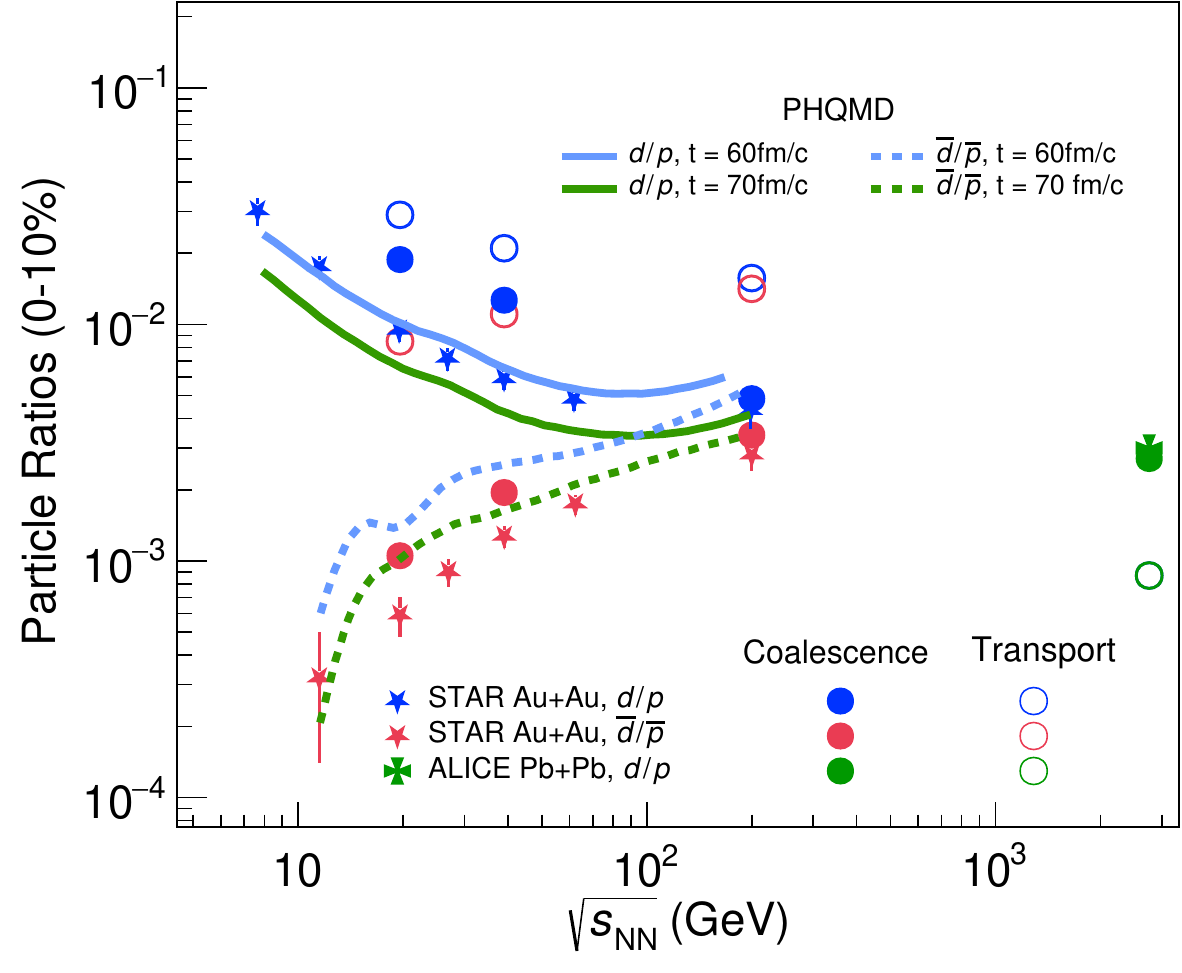}
    \caption{\justifying (Color Online) Particle ratios for deuterons over protons and anti-deuterons over anti-protons as a function of center of mass energies from \coal and \trpt modes. Solid lines are predictions from the PHQMD model~\cite{Glassel:2021rod}. Experimental measurement of $d/p$ and $\bar{d}/\bar{p}$ ratios from the STAR and ALICE experiments for the central Au+Au collisions at \sqsn = 19.6 GeV, 39 GeV, 200 GeV and central Pb+Pb collisions at \sqsn = 2.76 TeV are also shown~\cite{STAR:2019sjh,ALICE:2015wav}.}
    \label{ddbar_ppbar}
\end{figure}

The $d(\bar{d})$ over $p(\bar{p})$ ratios are presented in Figure~\ref{ddbar_ppbar} as a function of center of mass energy. Both modes capture the trends for $d/p$ and $\bar{d}/\bar{p}$ in Au+Au collisions. However, there are substantial quantitative discrepancies between the model predictions and the experimental data. In particular, the \trpt mode significantly overestimates the yield ratios compared to the experimental measurements, indicating a large deviation from the data~\cite{STAR:2019sjh}. The \coal mode overestimates the values, especially at \sqsn = 19.6 and 39 GeV, but gives a good description for \sqsn = 200 GeV and 2.76 TeV~\cite{ALICE:2015wav}. For completeness, we append additional model comparisons from~\cite{Glassel:2021rod} in Figure~\ref{ddbar_ppbar}. The predictions from the light nuclei cluster production with a PHQMD approach shows a comprehensive study with different cluster freezeout times or $``$physical'' times. The $d/p$ ratio is well described for t $ = 60$ fm/$c$ for Au+Au collisions. The $\bar{d}/\bar{p}$ ratios for t $ = 70$ fm/$c$ are closer to data and consistent to our \coal predictions. However, this is accredited to an over-predicted $\bar{d}$ yield by the model. \\

\begin{figure}[!ht]
    \centering
    \includegraphics[scale=0.45]{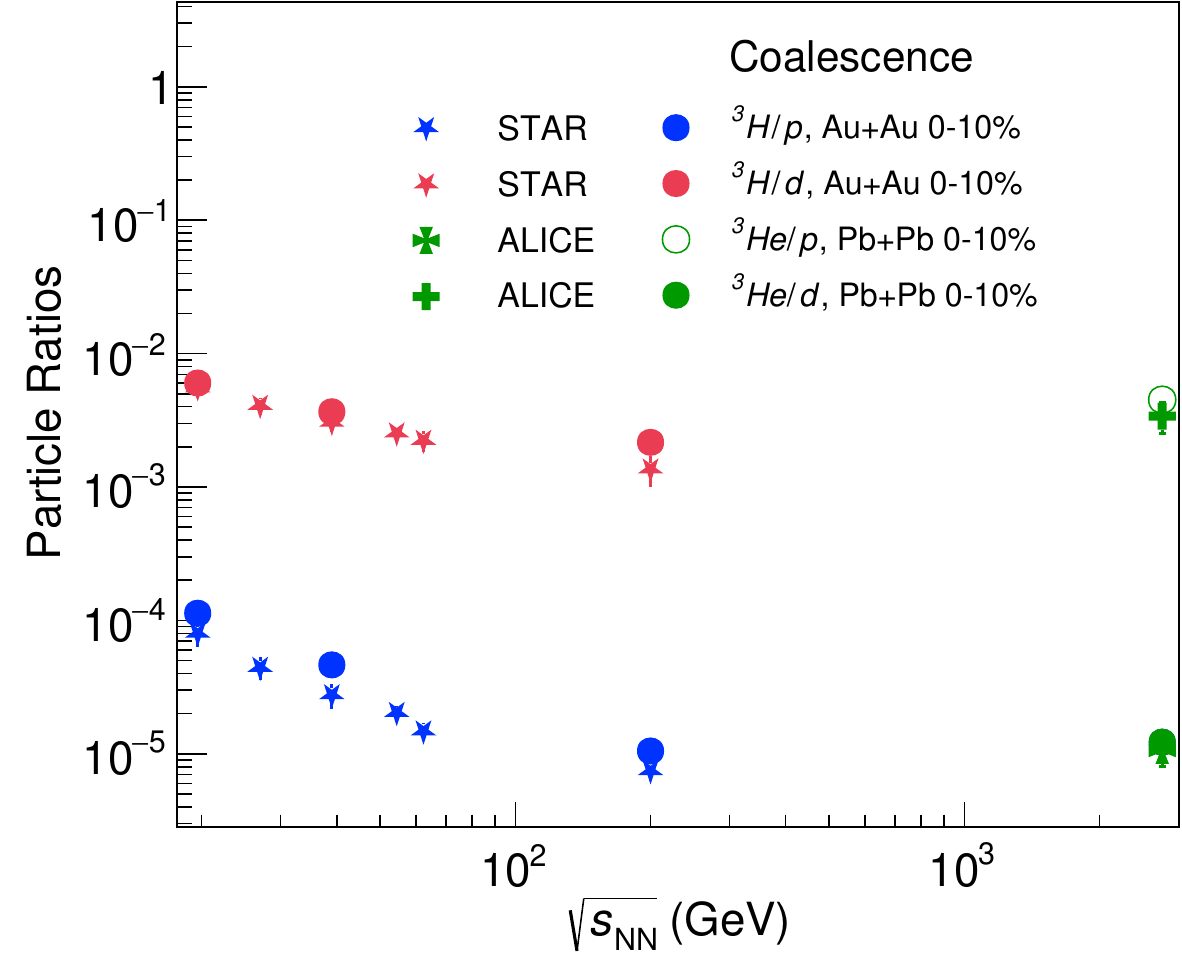}
    \caption{\justifying (Color Online) Particle ratios from the \coal mode for triton over proton and deuteron for 0-10\% central Au+Au and Pb+Pb collisions, helium-3 over deuteron for 0-10\% central Pb+Pb collisions and helium-3 over proton for 0-20\% central Pb+Pb collisions. Experimental measurement of $^3{H}/p$, $^3{H}/d$, $^3{He}/p$, $^3{He}/d$ ratios from the STAR and ALICE experiments for the central Au+Au collisions at \sqsn = 19.6 GeV, 39 GeV, 200 GeV and central Pb+Pb collisions at \sqsn = 2.76 TeV are also shown~\cite{STAR:2022hbp,ALICE:2015wav}.}
    \label{tHedp}
\end{figure}

In Figure~\ref{tHedp}, the $\rm{^3H}$, $\rm{^3He}$ \pt-integrated yields over $p$ and $d$ yields for central (0-10\%) Au+Au~\cite{STAR:2022hbp} and Pb+Pb collisions with \coal mode are presented~\cite{ALICE:2015wav}. All the ratios show the dependence with center of mass energy, and capture the trend of the experimental results. The $\rm{^3H}/p$ ratios are slightly overestimated for Au+Au collisions, but the $\rm{^3He}/p$ ratio for Pb+Pb collisions is well described. The $\rm{^3H}/d$ and $\rm{^3He}/d$ values at 0-10\% are also closely compatible with the experimental measurements~\cite{STAR:2022hbp,ALICE:2015wav}.\\

It can be concluded that while there are some discrepancies between the model's \pt-differential and \pt-integrated yields of light nuclei and experimental measurements from a quantitative standpoint, the model is still able to provide a reliable qualitative description of light nuclei yields as a function of \pt and center of mass energy. These findings demonstrate the effectiveness of a simple phase space coalescence approach in qualitatively describing light nuclei production across a wide range of collision energies and systems. The model's ability to offer a comprehensive explanation of experimental data with only a few tuning parameters is particularly noteworthy. Moving forward, these results can be further refined by leveraging the enhanced capabilities of the updated string melting AMPT model.\\

\begin{figure}[!ht]
    \centering
    \includegraphics[scale=0.33]{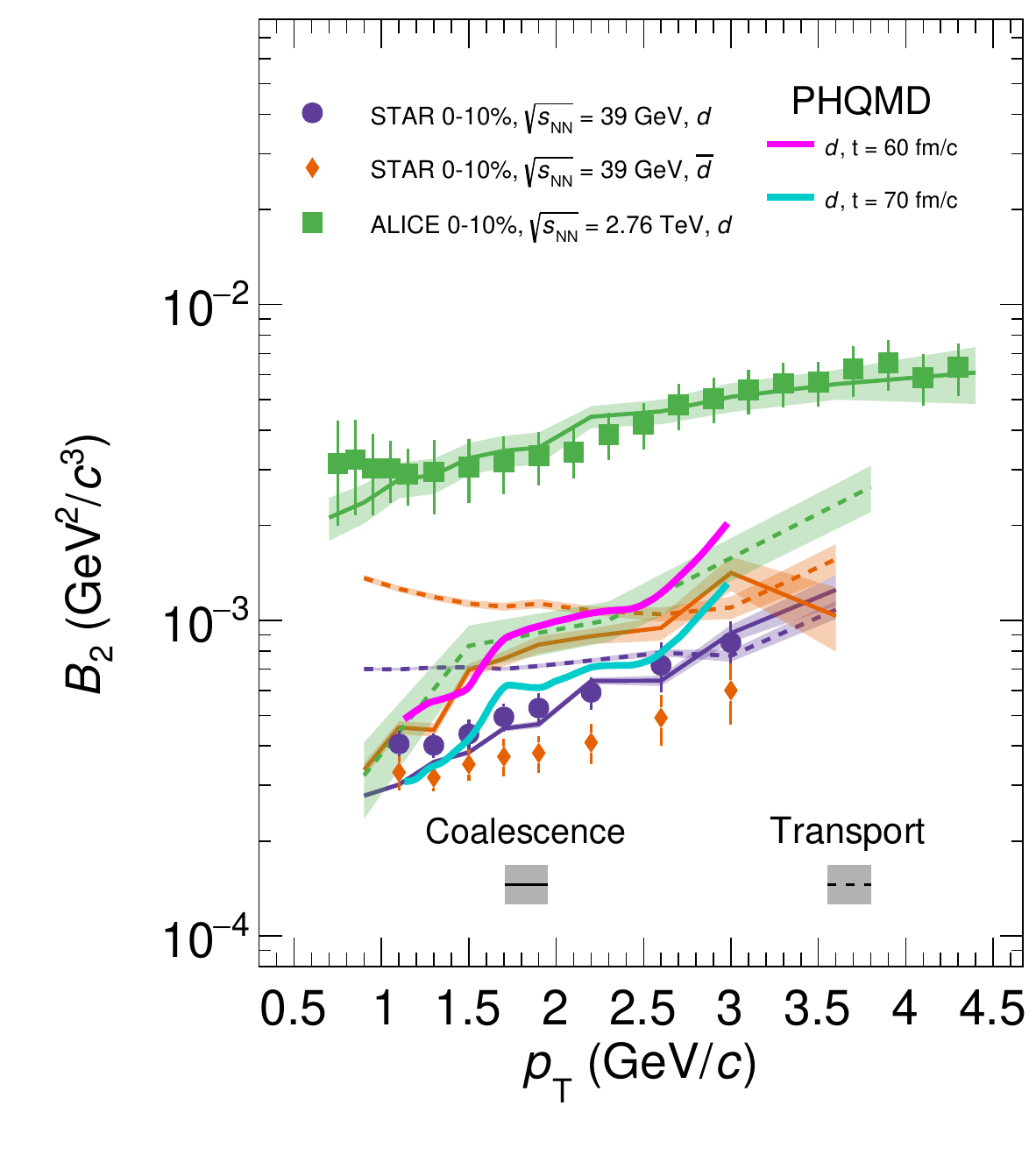}
    \includegraphics[scale=0.33]{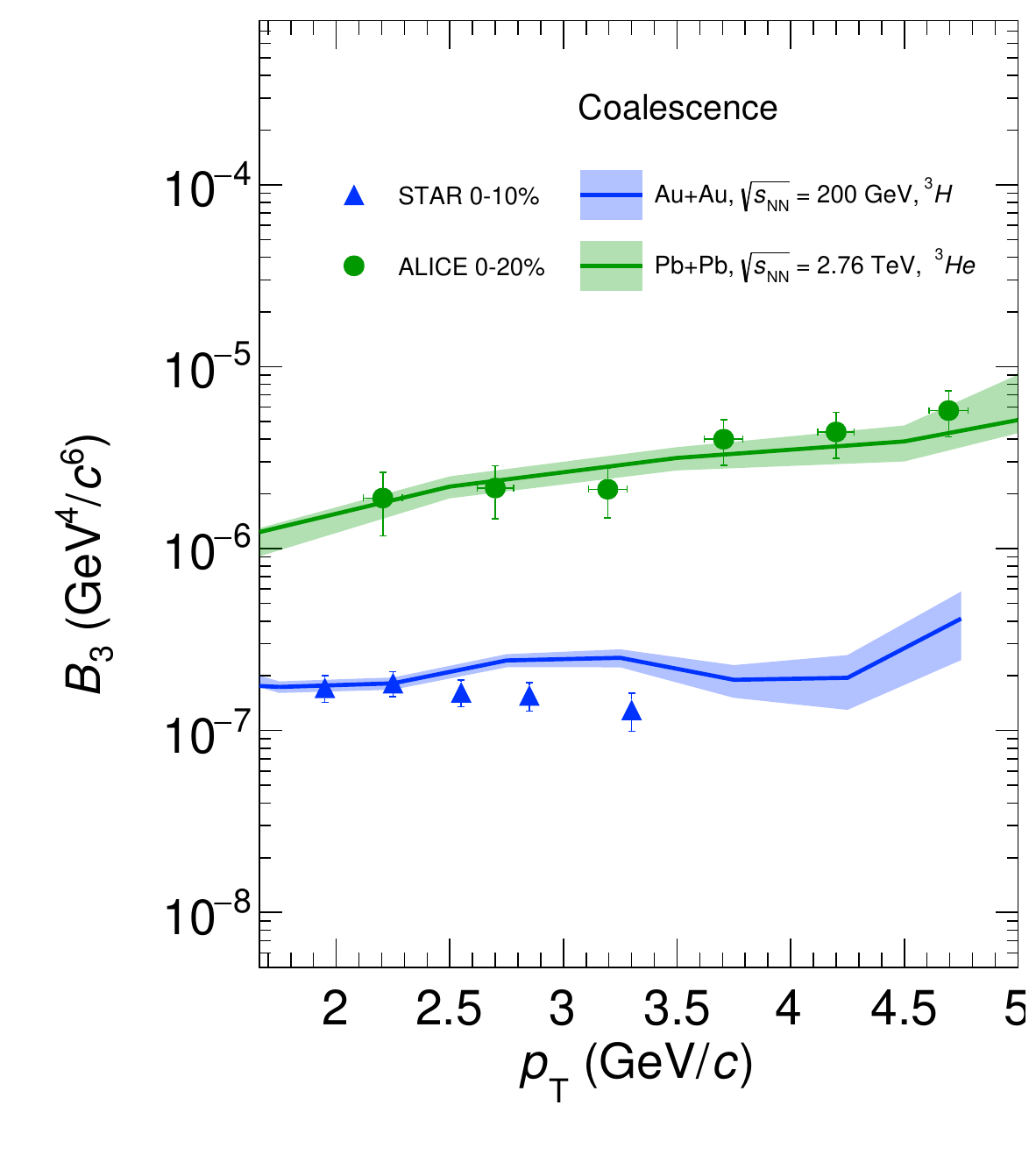}
    \caption{\justifying (Color Online) (Left) The $B_2$ parameter versus \pt of deuterons and anti-deuterons in central (0-10\%) Au+Au collisions at \sqsn = 39 GeV and central Pb+Pb collisions at \sqsn = 2.76 TeV from \coal (solid lines) and \trpt (dashed lines) modes. Predictions from the PHQMD model are superimposed~\cite{Glassel:2021rod}. (Right) The $B_3$ parameter versus \pt for triton and helium-3 in central (0-10\%) Au+Au collisions at \sqsn = 200 GeV and central (0-20\%) Pb+Pb collisions at \sqsn = 2.76 TeV from \coal (solid lines) mode. The results are compared to experimental measurements (symbols) from STAR and ALICE~\cite{STAR:2019sjh,Zhang:2020ewj,ALICE:2015wav}.}
    \label{b2200}
\end{figure}

The $B_{2}$ values as a function of \pt are calculated for central collisions at \sqsn = 39 GeV (Au+Au) and 2.76 TeV (Pb+Pb) using equation~\ref{b2formula}. Figure~\ref{b2200} (left) shows the results from \coal and \trpt modes compared to experimental measurements~\cite{STAR:2019sjh}\cite{ALICE:2015wav}. The \coal mode is promising in reproducing the increasing \pt dependence of deuteron $B_2$ for central Au+Au collisions at \sqsn = 39 GeV and 2.76 TeV. For deuterons, the model predictions are compatible with the experimental results for Au+Au as well as Pb+Pb collisions. The anti-deuterons at \sqsn = 39 GeV (Au+Au) shows an overestimated $B_2$ vs \pt with a finite slope. The \trpt mode is unable to reproduce the increasing dependence of $B_2$ with \pt for Au+Au collisions. Both deuteron and anti-deuteron results are largely overestimated by \trpt mode for \sqsn = 39 GeV (Au+Au). For Pb+Pb collisions, the \trpt mode is able to reproduce the increasing trend with \pt. However, the results are underestimated in comparison to the experimental measurements\cite{ALICE:2015wav}. The highlight of these results is the ability of the \coal mode to reproduce the increasing trend of $B_2$ with \pt. The PHQMD model results for deuteron $B_2$ for Au+Au collisions at \sqsn = 39 GeV also successfully reproduces the increasing trend with \pt, for cluster freezeout time t $= 70$ fm/$c$. A lower freezeout time overestimates the values. Moreover, the simple coalescence model outperforms in quantitative estimation of the $B_2$ for deuterons.\\

Furthermore, we calculate the $B_{3}$ values as a function of \pt for central collisions at \sqsn = 200 GeV (Au+Au) and 2.76 TeV (Pb+Pb), and we present the results from the \coal mode in Figure~\ref{b2200} (right). The predictions are fairly close to the experimental data for 0-20\% central Pb+Pb collisions~\cite{ALICE:2015wav} and 0-10\% central Au+Au collisions~\cite{Zhang:2020ewj}, respectively. The trend of $B_{3}$ as a function of \pt is quantitatively reproduced for helium-3s with good agreement with the experimental results. On the other hand, the data for triton $B_{3}$ shows a flat or decreasing dependence with \pt, whereas the model shows somewhat of an increasing trend.\\

\begin{figure}[!ht]
    \centering
    \includegraphics[scale=0.45]{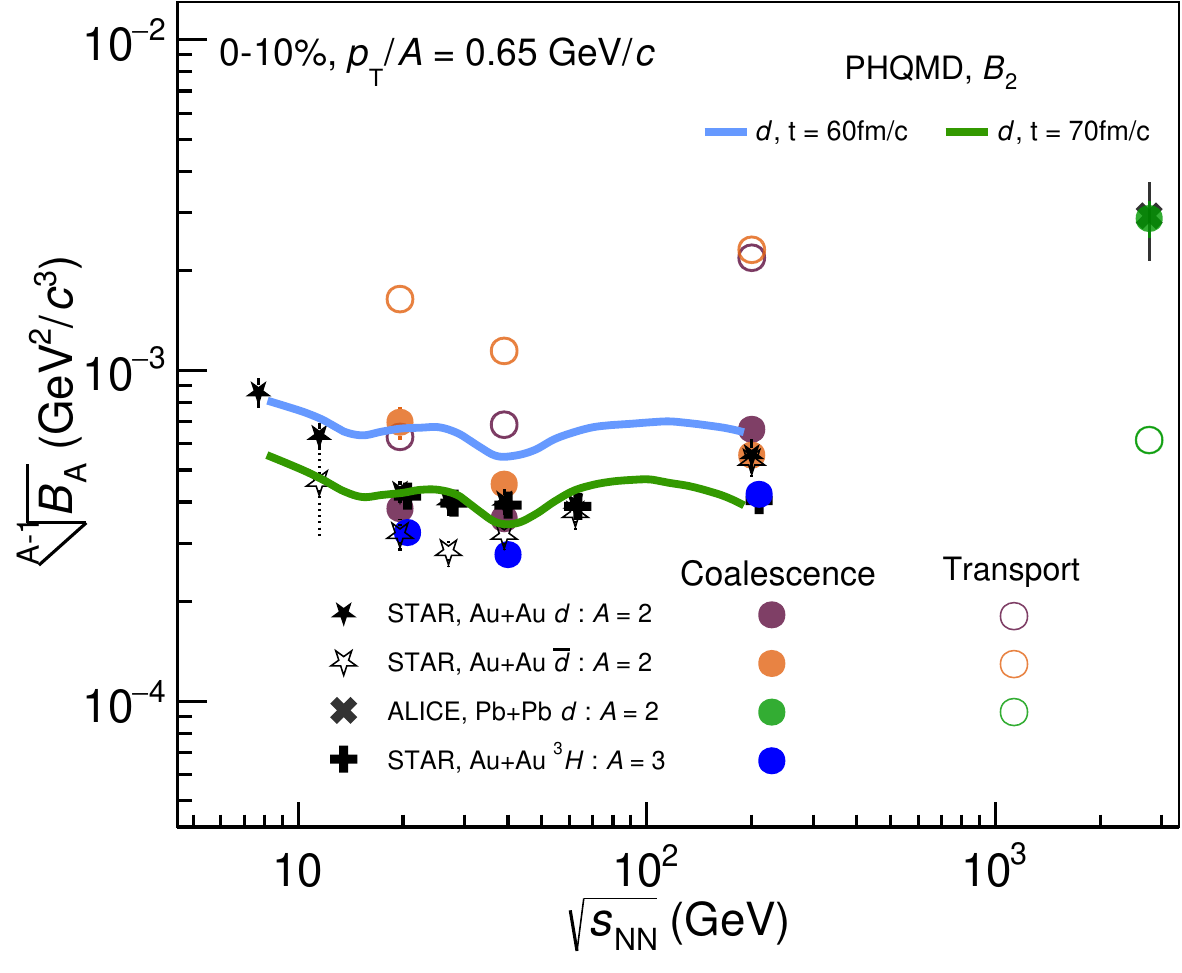}
    \caption{\justifying Coalescence parameter $B_2$ and
$\sqrt{B_3}$ as a function of \sqsn in central (0-10\%) Au+Au and Pb+Pb collisions for \coal and \trpt modes. Solid lines are predictions from the PHQMD model~\cite{Glassel:2021rod}. The results are compared with measurements from the STAR and ALICE experiments for $p_{\rm{T}}/A$ = 0.65 GeV/$c$. ~\cite{STAR:2019sjh,Zhang:2020ewj,ALICE:2015wav}.}
    \label{bAvssnn}
\end{figure}

Figure~\ref{bAvssnn} reports the $B_A$ as a function of center of mass energies for (anti-)deuterons, tritons and helium-3 for central Au+Au and Pb+Pb collisions. The values are obtained for $p_{\rm{T}}/A$ = 0.65 GeV/$c$, which is 1.3 GeV/$c$ for deuterons and 1.95 GeV/$c$ for triton and helium-3. The $B_2$ of deuterons from \coal shows a good match within uncertainties with the experimental results~\cite{STAR:2019sjh} across the \sqsn. The points for anti-deuterons at \sqsn = 19.6 and 39 GeV are systematically overestimated and appear higher than the $B_2$ for deuteron. For \sqsn = 200 GeV (Au+Au), the anti-deuteron $B_2$ falls within the uncertainties of the experimental results, nearly superimposed on the $B_2$ of the deuteron. The \trpt mode, overestimates the coalescence parameter for both deuteron and anti-deuteron for the Au+Au collisions. The same is underestimated for Pb+Pb collisions. The PHQMD predictions also show the deuteron $B_2$ as a function of collision energy for Au+Au collisions. Out of the two freezeout times, t $ = 70$ fm/$c$ shows a better agreement to the experimental measurements. The $B_3$ for tritons in Au+Au collisions are also shown and compared to experimental measurements from STAR~\cite{Zhang:2020ewj}. The results at \sqsn = 200 GeV is accurately described by the model, whereas at \sqsn = 19.6 and 39 GeV the values are slightly underestimated.

\section{Conclusions}

This article presents the results of a systematic study on the production of light nuclei in central Au+Au collisions at \sqsn = 19.6, 39, 200 GeV, as well as Pb+Pb collisions at \sqsn = 2.76 TeV using the AMPT driven coalescence afterburner model. The coalescence model successfully provides a comprehensive description of all light nuclei species across various collision energies in heavy-ion collisions. We also test the feasibility of (anti-)deuteron production processes implemented during the hadronic evolution in AMPT. Furthermore, we make a comparison of nuclei cluster production with a PHQMD approach with our calculations. The PHQMD approach considers deuteron production via potential interactions during the hadronic evolution. Overall, our phase space coalescence model out-performs the kinetic implementation of deuteron production in AMPT and performs closely with the PHQMD results.\\

The string melting AMPT model under the 
\coal mode struggles to quantitatively reproduce the experimental measurements of (anti-)proton \pt spectra for Au+Au collisions at \sqsn = 19.6, 39 GeV but provides a good description for Au+Au collisions at \sqsn = 200 GeV and Pb+Pb collisions at \sqsn = 2.76 TeV. The \trpt mode shows a better agreement of the (anti-)proton \pt spectra in all collision energies. While the predictions from \coal mode deviate from the data for the \pt differential yields of deuterons for Au+Au collisions at \sqsn = 19.6, 39 GeV, it gives a reasonable description for the deuterons in central Au+Au collisions at \sqsn = 200 GeV and Pb+Pb collisions at \sqsn = 2.76 TeV. A similar outcome is seen for tritons and helium-3s, with a slight deviation for Au+Au collisions at \sqsn = 19.6 and 39 GeV, with a good description at \sqsn = 200 GeV (Au+Au) and 2.76 TeV (Pb+Pb). The \trpt mode largely overestimate the deuteron and anti-deuteron yields as a function of \pt across all energies.\\

The \coal mode qualitatively describes the various particle ratios as a function of collision energy. The trend of the yields with center of mass energies is successfully described by the model predictions. The \coal results are fairly compatible with the  $d/p$, $\bar{d}/\bar{p}$ ratios predicted by the PHQMD model. The \trpt model overestimates these ratios. The $\bar{d}/d$ ratios are qualitatively reproduced by both \coal and \trpt mode for Au+Au collisions.\\

The coalescence parameter $B_2$ and $B_3$ as a function of \pt are estimated for RHIC and LHC energies. The results from \coal mode describe the trends of the coalescence parameter fairly well, reproducing the increasing dependence of $B_2$ with \pt from the experimental measurements. The results are better when compared to the PHQMD model for deuteron $B_2$. The $B_3$ vs \pt for helium-3 at \sqsn = 2.76 TeV (Pb+Pb) is well described by the model, however struggles to reproduce the trend for tritons at \sqsn = 200 GeV (Au+Au). The description of the \pt-spectra of (anti-)protons and light-nuclei by the model may vary at low and high \pt, but these effects are diminished for the predictions of $B_{A}$. The coalescence parameter $B_{A}$ as a function of \sqsn is also well described by the \coal mode for the deuterons for the RHIC and LHC energies. The \coal mode overestimates the anti-deuteron $B_{2}$ at \sqsn = 19.6 and 39 GeV, but sits well for \sqsn = 200 GeV. The $B_3$ values for triton and helium-3 are well reproduced for \sqsn = 200 GeV, and underestimates for \sqsn = 19.6 and 39 GeV.\\

In conclusion, the results show the effectiveness of a simple coalescence model and its ability to describe light nuclei production across different center of mass energies in heavy-ion collisions. We broadly observe that the model provides better predictions for higher collision energies (\sqsn = 200 GeV (Au+Au) and 2.76 TeV (Pb+Pb)). The model successfully covers various aspects of light nuclei production while accurately predicting a wide range of experimental observables. This study can be extended to hyper-nuclei production, as well as exotic states that are measured in heavy-ion experiments.

\section{Acknowledgements} 

 Author Y.B is thankful to Z.W. Lin for his helpful suggestions to this work, and to all the authors of AMPT. Y. B is grateful to Sudhir P. Rode for carefully reading the manuscript, and to Sumit Kundu for his contribution to the data generating process. Y.B is also thankful to Ravindra Singh and Swapnesh Khade for the fruitful discussions.
 





 \end{document}